\pdfoutput=1
\PassOptionsToPackage{
pdfencoding=auto,
pdfnewwindow=true,
pdfusetitle=false,
bookmarks=true,
bookmarksnumbered=true,
bookmarksopen=true,
pdfpagemode=UseThumbs,
bookmarksopenlevel=1,
pdfpagelabels=false, 
breaklinks=true,   
backref=false, 
colorlinks=true,
}{hyperref}
\RequirePackage{etex}
\PassOptionsToPackage{skip=0pt, font=small,labelfont=bf}{caption}
\documentclass[10pt,showpacs,letterpaper,aps,pra,twocolumn,nofootinbib,superscriptaddress,floatfix]{revtex4-2} 
\usepackage{hyperref}

\usepackage{amsthm,amsmath,amssymb,graphicx,comment,dsfont} 
\usepackage{centernot}
\usepackage[style=american]{csquotes}
\usepackage{color,soul}
\usepackage{bbold}
\usepackage[none]{hyphenat}
\usepackage{fancyhdr}
\newcommand{\beq}{\begin{equation}}
\newcommand{\eeq}{\end{equation}}
\newcommand{\bea}{\begin{align}}
\newcommand{\eea}{\end{align}}

\usepackage[capitalize]{cleveref}

\usepackage{graphicx}
\usepackage[usenames,dvipsnames,table]{xcolor} 
\usepackage{mathtools}
\usepackage[normalem]{ulem}
\input{preamble}

\definecolor{googleblue}{RGB}{34, 0, 204}
\definecolor{panblue}{RGB}{0,24,150}
\definecolor{carmine}{RGB}{150, 0, 24}
\hypersetup{
unicode=true,
bookmarksnumbered=false,
bookmarksopen=false,
breaklinks=false,
colorlinks=true,
linkcolor=carmine,
citecolor=googleblue,
urlcolor=panblue,
anchorcolor=OliveGreen}

\newcommand{\Hilb}[1][]{\ensuremath{\mathcal{H}_{#1}}}
\newcommand{\Ket}[1]{\ensuremath{\left \vert #1 \right \rangle}}

\newcommand{\BraKet}[2]{\ensuremath{\left \langle #1 \middle \vert #2 \right \rangle}}
\newcommand{\KetBra}[2]{\ensuremath{\left \vert #1 \middle \rangle \middle \langle #2 \right \vert}}
\newcommand{\Proj}[1]{\ensuremath{\KetBra{#1}{#1}}}
\allowdisplaybreaks

\usepackage{subcaption}
\usepackage[labelformat=simple]{subcaption}

\usepackage{microtype}
\microtypecontext{spacing=nonfrench}
\microtypesetup{
expansion={true,nocompatibility},
protrusion={true,nocompatibility},
activate={true,nocompatibility},
tracking=true,
kerning=true,
spacing={true}
}

\usepackage[all=normal,floats=tight,mathspacing=tight,wordspacing=tight,paragraphs=normal,tracking=tight,charwidths=tight,mathdisplays=normal,sections=normal,margins=normal]{savetrees}

\everypar=\expandafter{\the\everypar\loosness=-1 }
\newcommand{\nocontentsline}[3]{}
\let\oldaddcontentsline\addcontentsline
\newcommand{\tocless}[2]{%
  \let\addcontentsline=\nocontentsline#1{#2}
  \let\addcontentsline\oldaddcontentsline}

\begin{document}
\title{Copenhagenish interpretations of quantum mechanics}
\author{David Schmid}
\email{davidschmid10@gmail.com}
\affiliation{Perimeter Institute for Theoretical Physics, Waterloo, Ontario, Canada, N2L 2Y5}
\author{Y\`{i}l\`{e} Y{\=\i}ng}%for arxiv: Y\`il\`e Y\=ing
\email{yying@perimeterinstitute.ca}
\affiliation{Perimeter Institute for Theoretical Physics, Waterloo, Ontario, Canada, N2L 2Y5}
\affiliation{Department of Physics and Astronomy, University of Waterloo, Waterloo, Ontario, Canada, N2L 3G1}
\author{Matthew S. Leifer}
\email{leifer@chapman.edu}
\affiliation{Institute for Quantum Studies \& Schmid College of Science and Technology, Chapman University, One University Drive, Orange, CA, 92866, USA}

\date{\today}

\begin{abstract}
We define a class of \emph{Copenhagenish interpretations} encompassing modern interpretations that follow the Copenhagen spirit. These interpretations are characterized by four postulates: Observers Observe, Universality, Anti-$\psi$-ontology, and Completeness. We explain why such interpretations are not equivalent to the textbook (or orthodox) interpretation, nor to the view that one should shut up and calculate, nor to strict operationalism. We then discuss what lessons are implied for Copenhagenish interpretations by the measurement problem, the Wigner's friend thought experiment, and the simple variants of the Wigner's friend thought experiment that we term Wigner's enemy, stalkee, and penpal. In particular, we discuss how Copenhagenish interpretations give multiple distinct descriptions of each experiment, where these descriptions are each individually true, yet cannot be combined into any single description. To make such interpretations consistent, then, one requires epistemological constraints forbidding certain perspectives from being combined. We discuss these constraints, their motivations, and some of the challenges they introduce.
\end{abstract}

\maketitle
\tableofcontents

\section{Introduction}

More than a century after the inception of quantum mechanics, the Copenhagen interpretation remains popular among physicists.
However, the very meaning of the term ``Copenhagen interpretation'' is debated among scholars of the foundations and interpretations of quantum mechanics.  This is because the views of the Copenhagen founders differed in significant ways, are self-contradictory even in different writings by the same author, and can be difficult to interpret (especially the writings of Bohr)\footnote{Howard ~\cite{Howard2004} argues that the idea of a single ``Copenhagen Interpretation'' is largely an invention of Heisenberg in the 1950's.  Before that time, people just spoke of ideas ``in the Copenhagen spirit''.}.

Several more coherent modern interpretations of quantum mechanics draw inspiration from the Copenhagen interpretation.  Examples include Bub and Pitowsky's ``information'' interpretation~\cite{Bub2005, Bub2010, Bub2016}, Healey's quantum pragmatism~\cite{Healey2012,  Healey2017, Healey_2018, Healey2020}, Rovelli's relational quantum mechanics~\cite{rovelli1996relational, Rovelli2018}, the QBism of Fuchs, Schack, et. al.~\cite{FuchsMerminSchack,Fuchs2016}, Zwirn's ``Convivial Solipsism'' \cite{Zwirn2016, Zwirn2024}, and the views of various others, such as Peres \cite{Peres1978, peresQuantum1995}, and Brukner and Zeilinger \cite{Brukner2003, Brukner2005}.  These interpretations are more consistent and have been worked out in more detail than the original Copenhagen view. We call them Copenhagen\emph{ish} interpretations\footnote{Another common name is \emph{neo-Copenhagen} interpretations, but it is difficult to find a terminology that all proponents can agree upon.  Fuchs objects to ``neo-Copenhagen'' on the grounds that ``Copenhagen'' is most closely associated to the views of Bohr, whereas QBism draws more strongly from Pauli and Wheeler~\cite{Fuchs2011, Fuchs2017}.  Thus, the \emph{ish} in Copenhagenish is meant to indicate a vague family resemblance rather than a direct line of influence.}.

To date, there has been no precise definition of a Copenhagenish interpretation, which makes it difficult to understand the common principles underlying them.  Indeed, Copenhagenishts are often more keen to emphasize what \emph{distinguishes} their view from other Copenhagenish interpretations, so that they can be seen to be contributing something new to the debate. Meanwhile, critics may find it convenient to dismiss all variants of Copenhagenishism by pointing at the impenetrability and incoherence of certain specific versions thereof, particularly those due to Bohr.  In this work, however, we aim to present Copenhagenish interpretations in as charitable and robust a way as possible---to identify sensible core principles that underpin them, and to attempt to make these principles consistent. 
This can also be useful for distinguishing which criticisms of Copenhagenish ideas attack the core premises from those that aim to discriminate among specific variants, and can help us understand why certain features of specific variants might be desirable.

In this work, we propose a precise definition of what constitutes a Copenhagenish interpretation, based on four postulates: 
\begin{enumerate}
\item Observers Observe,
\item Universality,
\item Anti-$\psi$-ontology,
\item Completeness.
\end{enumerate}
Roughly, these postulates state the following. Observers Observe states that sufficiently complex agents (such as humans) observe definite outcomes when making quantum measurements. Universality states that quantum theory (including the unitary description of dynamics, the Born rule and the measurement postulates) can be applied to arbitrary subsystems of the universe. Anti-$\psi$-ontology states that the wavefunction is not an intrinsic property of an individual quantum system. Finally, Completeness states that there is no deeper description of reality beyond the quantum formalism. 

We believe that this characterization should be acceptable to most proponents of these interpretations. Although we do not attempt a detailed historical analysis\footnote{We will not engage in the common practice of quoting Bohr at length to try to argue that his interpretation is Copenhagenish in our sense.  That is a job for historians.  Indeed, several of the issues currently regarded as central to the foundations of quantum theory, such as the nature of the quantum state, were not formulated precisely at the time of Bohr's writing, so you will not find precise statements on them.  It is a fool's errand to try to pin down his views on such topics.  Much more interesting is to argue the merits of contemporary Copenhagenish interpretations that are in a similar spirit, but have been worked out to a greater degree of clarity and coherence.}, our view is that the original Copenhagen interpretation fits into this class, as do the examples of Copenhagenish interpretations given earlier, as well as the views of most physicists who say that they support ``the Copenhagen interpretation''. 

Copenhagenish interpretations are often described as antirealist, positivist, empiricist, instrumentalist, operationalist, etc., 
(in this paper, we use these words interchangeably), usually by those who intend these terms as insults. Although we do not believe these terms are strictly accurate, it is true that Copenhagenish interpretations attribute few, if any, intrinsic ontological properties to systems being described according to quantum mechanics.  This allows them to easily evade the measurement problem~\cite{maudlin1995three} and no-go theorems such as Bell's theorem~\cite{Bell,wood2015lesson} or noncontextuality no-go theorems~\cite{unphysical}. 

Indeed, many no-go theorems about quantum theory start with an assumption like ``assume every system has some objective physical properties, $\lambda$, that exist independently of the observer''.  This renders the no-go theorems irrelevant to Copenhangenish interpretations, as Copenhagenishts either deny that such $\lambda$'s exist or deny that there is any meaningful way of describing them. However, one class of  arguments that do not require assumptions of this sort are the Wigner's friend thought experiment~\cite{Wigner1995} and modern extensions of it~\cite{schmid2023review,brukner2015quantum,Frauchiger2018,bongStrong2020,Cavalcanti2021}. 

In his famous thought experiment~\cite{Wigner1995}, Wigner considered what happens when an observer---his ``friend''---makes a quantum measurement and, some time later, Wigner asks the friend what he observed.  This can be analyzed from two perspectives.  Before Wigner interacts with them, Wigner can model the system and friend as two quantum systems interacting unitarily, so that they end up in an entangled superposition state.  For Wigner, the collapse to a definite measurement outcome supposedly only happens later, when he interacts with his friend.  However, from the friend's perspective, the collapse must have occurred earlier when they actually made the measurement and presumably saw the outcome.  Wigner argued that these two perspectives are contradictory.  This is because he analyzed the experiment according to the \emph{orthodox} interpretation (also called the \emph{textbook} or \emph{Dirac-von Neumann} interpretation) of quantum mechanics.  In this interpretation, the two descriptions prescribe different objective physical properties to the system, and so are contradictory.  Not being a solipsist, Wigner believed that his friend does indeed experience a definite outcome, so he argued that only the friend's description is correct and that the quantum state must objectively collapse by the time a measurement outcome enters the awareness of a conscious agent\footnote{Wigner used this as an argument against the materialist view of consciousness, and in favor of the view that consciousness has an effect on matter.  He was not particularly concerned with the problems of measurement in quantum mechanics in the paper that introduced his friend~\cite{Wigner1995}, although modern treatments tend to view the Wigner's Friend experiment as a refinement of Schr{\"o}dinger's cat or the measurement problem.}.

However, as we explain in more detail later, the orthodox interpretation is not Copenhagenish.  Copenhagenish interpretations assign far fewer intrinsic objective properties to quantum systems than the orthodox interpretation, and as a result the two descriptions arising in Wigner's argument can coexist without contradiction.  

Consequently, Wigner's friend arguments (unlike many other no-go theorems in quantum foundations) can more meaningfully be used to obtain insights into the Copenhagenish picture of the world. In this work, we give a Copenhagenish analysis of a variety of simple Wigner's friend arguments, including the original Wigner's friend thought experiment and three variants that we term Wigner's  enemy, Wigner's stalkee, and Wigner's pen pal. These analyses serve as a basis for  the more difficult task of providing a detailed Copenhagenish analysis of more complicated thought experiments, such as Extended Wigner's Friend scenarios \cite{schmid2023review,brukner2015quantum,Frauchiger2018,bongStrong2020,Cavalcanti2021} (a task we leave to future work).

A key lesson illustrated by these simple thought experiments is that Copenhagenish interpretations necessarily provide multiple complementary descriptions of an experiment, where each description is valid, but where one cannot combine the different descriptions into a single description. Any given agent may adopt any given description, but no agent may combine two incompatible descriptions. For consistency, this implies that Copenhagenish interpretations must impose epistemological constraints on how one combines certain types of information. Our work elucidates the need for these constraints as well as the difficulties they introduce.

\section{Copenhagenish interpretations}
\label{Copenhagen}

Broadly speaking, a \emph{Copenhagenish} interpretation is one that bears a family resemblance to the ideas of the Copenhagen founders.  In particular, Copenhagenishts want to view quantum theory as a fundamental physical theory, potentially applicable to any system in the universe, and in no need of ``completion'' with additional postulates or variables, while at the same time denying that the quantum formalism supplies a literal description of the intrinsic properties of a quantum system.  It might seem that the measurement problem~\cite{maudlin1995three}, Schr{\"o}dinger's cat~\cite{schroedingerGegenwaertige1935}, Wigner's friend~\cite{Wigner1995}, and related thought experiments already pose problems for such interpretations, but we argue that they do not.  These thought experiments show that there are situations in which different observers will assign different quantum states to the same quantum system, but this need not be a problem if the quantum state is not an intrinsic property of an individual quantum system, but rather represents something more like the observer's knowledge of a quantum system.  Just as different observers may assign different probability distributions to the same classical system if they have access to different information, the same can be true of quantum state assignments.  With this in mind, we now explain the four defining postulates of Copenhagenish interpretations.

\subsection{Four defining postulates}

The four defining postulates are:
\begin{enumerate}
	\item Observers Observe 
	\item Universality
	\item Anti-$\psi$-ontology
	\item Completeness
\end{enumerate}
We explain each of these in detail in the following sections.

\subsubsection{Observers Observe}
\label{Copenhagen:OO}

Observers Observe says that there exist many entities in the universe, or at least on planet Earth, that experience definite outcomes when they make quantum measurements.  I know, from personal experience, that I am one such entity\footnote{The authors are all theorists, so it has been a long time, but most of us have taken laboratory classes as undergraduates.}.  Assuming that solipsism is incorrect, I infer that other similarly constituted entities, such as students in a quantum mechanics lab, also count as observers who have definite experiences when they make quantum measurements.  

Note that this postulate does not state that human observers, consciousness, or anything like that, is \emph{required} for something to count as an observation.  One could, for example, accept an account of when definite outcomes occur based on environmental decoherence.  Cats, robots, sufficiently complicated experimental equipment, or even (in some views \cite{rovelli1996relational, Rovelli2018})
any physical system at all, could well be capable of registering a definite outcome in a quantum experiment.  The exact account of the nature of observation will, of course, depend on the precise Copenhagenish interpretation under consideration.  The point is just that, whatever else counts as an observation, we know from experience that we ourselves see definite outcomes and we infer, by assuming no solipsism, that other human observers do too.  Human observation is thus a \emph{sufficient} condition for a definite experience, but it need not be necessary.

Although Observers Observe says that observers experience definite outcomes, it does not specify whether the outcomes exist objectively for everyone, or just for the observer who makes the measurement.  Copenhagenish interpretations fall into two categories depending on their stance on this question.
 
\begin{enumerate}
	\item Objective:  When you make a measurement and observe the result, then the result that you have obtained is an objective fact about the universe.  Other observers may not have access to this objective fact, and the subsequent evolution of the universe may make you forget this fact and/or make it fundamentally impossible for other observes to acquire information about this fact, but it is an objective fact nonetheless. 
	\item Relativist\footnote{In earlier talks on this topic, we used the term ``perspectival'' rather than ``relativist''.  We did this because, in philosophy of science, ``relativism'' is often associated with extreme views such as social constructivism about scientific theories.  To avoid raising that spectre, we used ``perspectivism''.  However, perspectivism is most commonly understood in epistemic terms -- the apparent objectivity of knowledge is an illusion and there is no perspective-free view from which we can ascertain the nature of reality in itself.  \emph{All} Copenhagenish interpretations are perspectival in this sense, so this term cannot be used to distinguish two classes of Copenhagenish interpretation.  Relativism is really the correct term to use, as it is broad enough to encompass relative notions of truth as well as relative notions of knowledge.  To be precise, in the terminology of~\cite{Baghramian2025}, the notion of relativism we need is strong, local, alethic relativism about the results of quantum measurements.  ``Strong'' means that what is true for one observer may be false for another, not just inaccessible or undefinable.  ``Local'' means that we are being relativist about one specific thing, in this case the outcomes of quantum measurements in situations where observers disagree about where the Heisenberg cut must be placed, and not wholesale about everything.  ``Alethic'' means that we are being relativist about truth.  Perspectivism is also a variety of relativism, about knowledge rather than truth, but we need the stronger notion here.  The relativism we need is extreme, in that it is about truth, but it is also limited in scope.  It does not imply global relativism and, in particular, it does not imply cultural relativism of any sort.}: When you make a measurement and observe the result, then the result you register is an objective fact \emph{for you}.  There is no fact of the matter about the result \emph{for me} unless and until I repeat the measurement myself on the same system or interact with you, e.g.,  you tell me the result you obtained.  
\end{enumerate}

Note that the objective version of Observers Observe is essentially the same as the Absoluteness of Observed Events (AOE) assumption that has been used in several extended Wigner's friend arguments~\cite{schmid2023review,brukner2015quantum,bongStrong2020}. Of the Copenhagenish interpretations mentioned above, Rovelli's relational quantum mechanics and QBism are clearly relativist, and so reject this assumption.  

The objective view is in line with straightforward empiricism, in which our observations (the only things to which we have reliable direct access) are the foundation on which science is built.   Relativist Copenhagenish interpretations threaten to undermine this view, because they render scientific observations subjective.  Although we reject the idea that any Copenhagenish interpretation is wholesale empiricism, objective Copenhagenish interpretations are empiricist in this one basic aspect, and this has been part of the Copenhagen view from the beginning.  
Heisenberg arrived at matrix mechanics by focusing on the relations between observable quantities, such as atomic emission spectra, instead of trying to construct a mechanical explanation based on potentially unobservable entities.  The objectivity of these observed phenomena is not under question in objective Copenhagenish interpretations---they are the very bedrock on which the interpretation is built.  This lead Heisenberg and Bohr to incorporate this aspect of empiricism into their view.  

\begin{quotation}
It seems sensible to discard all hope of observing hitherto unobservable quantities, such as the position and period of the electron... Instead it seems more reasonable to try to establish a theoretical quantum mechanics, analogous to classical mechanics, but in which only relations between observable quantities occur.---Heisenberg~\cite{heisenberg1925Uber}
\end{quotation}

\begin{quotation}
\ldots all unambiguous information concerning atomic objects is derived from the permanent marks such as a spot on a photographic plate, caused by the impact of an electron left on the bodies which define the experimental conditions.---Bohr~\cite{Bohr1963}
\end{quotation}

This empiricism has also survived into modern objective Copenhagenish interpretations:

\begin{quotation}
Consistency requires that the result of observing the detectors by
another instrument is the same as if the detectors themselves are considered as
the ultimate instrument. This is what is meant by the claim that there is an
\emph{objective} record of the experiment. The role of physics is to study relationships
between these objective records. Some people prefer to use the word ``inter-subjectivity,'' which means that all observers agree about the outcome of any
particular experiment.---Peres~\cite{peresQuantum1995}
\end{quotation}

In contrast, relativist Copenhagenish interpretations are much more radical.  Depending on the specific interpretation, a measurement and its outcome might only exist relative to: an observer or agent~\cite{FuchsMerminSchack,fuchs2013quantum}, another interacting physical system~\cite{rovelli1996relational, Rovelli2018,dibiagioStable2021}, the placement of the Heisenberg cut\footnote{We introduce the Heisenberg cut in  Section~\ref{sec:consequence}.  In our view, it is not clear whether one can meaningfully take measurement outcomes to be defined only relative to an agent's placement of the Heisenberg cut, given that the cut is a construct for reasoning rather than a physical entity. See also the discussion in Section VIII.D of Ref.~\cite{schmid2023review}.}, or which measurements will be made in the future~\cite{renes2021consistency}.  If measurement outcomes only have meaning from one of these perspectives, then combining statements about {\em different} observers' perspectives into a consistent global picture will not always be possible.

Perhaps one of the main reasons why so many physicists pay lip service to the Copenhagen interpretation is because they approve of the apparent hard headed empiricism of the objective Copenhagenish view.  They do not want to get derailed into long discussions about interpretation.  If they only have to focus on regularities in observed phenomena then they can just get on with their work.  From this point of view, relativist Copenhagenish interpretations are shocking because they threaten to undermine the existence of objective experimental records on which empirical reasoning is based.  

In a relativist interpretation, there need be no single, global perspective that is consistent with all of the individual perspectives; rather, the world is comprised of a set of disconnected realities. This sort of far fetched conclusion is precisely what many physicists are trying to avoid by advocating Copenhagen. But as we will see when analyzing various Wigner's friend thought experiments, objective Copenhagenish interpretations lead to constraints on epistemology that have no obvious justification (apart from being required for internal consistency of the interpretation), whereas relativist Copenhagenish interpretations make these constraints arise naturally as a consequence of the (admittedly radical) metaphysics they support.

\subsubsection{Universality} \label{Copenhagen:U}

Universality states that quantum theory is a fundamental physical theory, and that anything in the universe (if perhaps not everything at the same time) can in principle be described by quantum theory.  In other words, there are no fundamentally ``classical'' or ``nonquantum'' systems in the universe. The key components of this universal description are the unitary dynamics, the Born rule for computing the probabilities of measurement outcomes, and the state-update rules upon measurement. 

Note that we do not regard the projection postulate as an axiom of quantum theory.  Whether it holds depends on how the measurement is implemented; e.g., it does not hold when the system is absorbed by the measuring device. Rather, we take the ``measurement postulates'' to mean the Born rule together with the necessary change to the quantum state upon learning the measurement outcome, as described by a \emph{quantum instrument}~\cite{NielsenAndChuang}. The projection postulate is just one example of a quantum instrument, and while projective measurements can be implemented in such a way that it holds, it is not the most general possibility. 

Provided that the system of interest is sufficiently isolated, Universality implies that its evolution can always be described unitarily. In particular, this applies to macroscopic systems (such as measuring devices and observers) in addition to microscopic systems (such as atoms).  This is in conflict with collapse theories, which postulate a dynamics that is fundamentally nonunitary in order to effect an \emph{objective} collapse of the quantum state. However, the Copenhagenish position on Universality {\em also} differs from ``unitary-only'' interpretations~\cite{kastner2020unitary} that reject the fundamental status of the measurement postulates, such as Everett/many-worlds and the de Broglie-Bohm theory.
In a Copenhagenish view, unitary dynamics and the measurement postulates have equal status.
%In a Copenhagenish view, not only is unitarity applicable to any system being described quantum mechanically, but the measurement postulates are {\em also} fundamentally applicable.

There is a certain tension in this---and, indeed, between Universality and Observers Observe.  In a measurement interaction, the joint system consisting of the system to be measured and observer\footnote{We typically define the ``observer'' to include not just the human observer, but also the measurement device and as much of the environment as necessary such that its interaction with the system can be described by unitary dynamics.  If there is environmental decoherence, then the part of the environment that is involved in the decoherence process is included as part of the ``observer''.} is an isolated system, and so their interaction should be describable unitarily. On the other hand, since Observers Observe, the observer must see a definite measurement outcome, so the post-measurement state would be given by the measurement postulates. On the face of it, these two descriptions seem contradictory, as they lead to two different quantum states. However, as we shall see in the next section, the Anti-$\psi$-ontology postulate says that quantum states do not represent intrinsic properties of a system, so two different ways of updating the state, leading to two different state assignments, do not necessarily represent contradictory statements about reality.  All that matters is whether they agree on the probabilities for the outcomes of measurements that are actually performed.  

Modern Copenhagenish interpretations are normally quite explicit about Universality.  As one example, Brukner states~\cite{brukner2015quantum}
\begin{quotation}
The measurement instrument and the observer can be included in the quantum mechanical description, and then observed by someone else, a ``superobserver'', for whom the original measurement instrument loses its previous status as a means for acquiring knowledge.
\end{quotation}
Here, Brukner is explicit that observers and measuring devices can either be described quantum mechanically or as classical systems, and that the two descriptions serve different purposes.
Rovelli states that ``\emph{any} physical system can play the role of Copenhagen's `observer' ''~\cite{Rovelli2018}.  Universality is implicit in this because the interaction between two microscopic systems would normally be described unitarily, so Rovelli is stating that this should not be treated differently from the interaction of a system with a measuring device.  In both cases, there is unitary evolution \emph{and} the realization of a definite outcome. Here, contradiction is avoided by making all physical properties relative to other physical systems, i.e.\ the measurement has an outcome relative to the measuring device, but it does not have an outcome relative to an external system that has not interacted with the system or measuring device.  From the perspective of this latter system, the interaction is described unitarily.

Although universality states that any part of the universe can be described according to quantum mechanics by some observer, it does not mean that there is a meaningful description in which \emph{every} part of the universe is described according to quantum mechanics.  
In particular, as we shall explain, the definite outcomes experienced by observers cannot be described within the quantum formalism in a Copenhagenish interpretation.  Hence, if you want to reason about the outcomes of some given observer, then that observer must be described classically\footnote{By classically, we just mean that the Heisenberg cut is placed before the observer, so that they are not included in the quantum description and the outcomes of their observations  can be treated as classical random variables.  Contra Bohr, we do not mean that they must be given a description terms of the quantities of classical mechanics, e.g., position and momentum.}. This is so even if you think that the observer, system, and measuring device together form an isolated system which you would otherwise describe as a composite quantum system evolving unitarily.

Despite appearances, this is consistent with observers being quantum systems and there being no limit on what can be described according to quantum mechanics.  \emph{I} can describe \emph{you} according to quantum mechanics and \emph{you} can describe \emph{me} according to quantum mechanics, but, if I want to reason about my own measurement outcomes, \emph{I} am not allowed to describe \emph{myself} according to quantum mechanics.  Therefore, you can describe me together with the systems I am measuring via unitary evolution provided you do not want to reason about my measurement results. The reason why I cannot describe myself according to quantum mechanics in such a case is not that I am not a quantum mechanical system, but rather because I have chosen to analyze my observations, and these can only be described in classical terms. And if {\em you} wish to reason about my outcomes, then you too must describe me classically, and for the same reasons. Thus any valid description can be adopted by any particular observer with the relevant information, as one would expect. 

In complex enough experiments with multiple observers, it can be difficult to see how the different descriptions one can give of the experiment all fit together. In order to maintain consistency, it will become clear that the different descriptions generally cannot be combined into a single perspective. We will return to these considerations repeatedly in the coming sections (perhaps most explicitly in Section~\ref{sec:consequence}).

\subsubsection{Anti-$\psi$-ontology}

\label{Copenhagen:APO}

Anti-$\psi$-ontology is the view that the quantum state $\Ket{\psi}$ is not an intrinsic property of an individual quantum system\footnote{In the language of ontological models, it is not an ontic state.}. This has to be formulated carefully, because Copenhagenishts often disagree about what the quantum state \emph{does} represent.  It may represent knowledge \cite{peresQuantum1995}, information \cite{Bub2005, bub2008dogmas}, beliefs  \cite{fuchs2009quantumbayesian, Fuchs2016}, or advice \cite{Healey2012, Healey2017}, about a quantum system, where ``about a quantum system'' can mean ``regarding the outcomes of measurements made on the system'', ``about responses to your interventions made on the system'', or other formulations of actions and responses.  The main point is that it does not mean ``about pre-existing intrinsic properties of the system''.

This is to be contrasted with the $\psi$-ontic~\cite{Harrigan} view, in which the quantum state of an individual quantum system is physically real.  For example, in de Broglie-Bohm theory, quantum particles have positions, but the quantum state also exists physically, as it determines how the positions of the particles evolve in each individual run of an experiment.  In Everett/many-worlds, the quantum state of the universe is not only physically real, it is moreover the \emph{only} fundamental ontology.  All appearances are to be explained in terms of it and its unitary evolution\footnote{This is a $\psi$-complete interpretation, which is a subspecies of $\psi$-ontic interpretations in which the quantum state is the \emph{only} ontology.}.

Operationally, if Alice prepares a system in the quantum state $\Ket{\psi}$ and hands it to Bob, who knows nothing about Alice's preparation procedure, then Bob would not be able to determine $\Ket{\psi}$ by making a quantum measurement on the system.  That is a prediction of quantum mechanics, on which all interpretations agree.  However, suppose that instead of just being given the quantum system, Bob somehow learns the complete description of all the intrinsic physical properties of the system.  In a $\psi$-ontic interpretation he would thereby be able to infer $\Ket{\psi}$, but in an Anti-$\psi$-ontic interpretation he would not.  He would need something like an ensemble of independent and identically prepared systems in order to gain enough information to reliably estimate $\Ket{\psi}$. 

The term $\psi$-epistemic is often used for the Anti-$\psi$-ontic view, but that is too narrow for our purposes because it singles out ``knowledge'' as the key term (whereas, as noted above, different Copenhagenish interpretations might prefer to focus on information, beliefs, advice, etc).  It is too narrow also because the term $\psi$-epistemic is defined in the framework of ontological models~\cite{Harrigan}, in which the quantum state represents knowledge about some underlying physical reality.  However, in a Copenhagenish interpretation in which $\Ket{\psi}$ represents knowledge, that knowledge is about the system's response to actions that may be performed upon it.

Copenhagenish interpretations differ on whether there is an objectively correct quantum state to assign to a system, given the available information about the system.  For example, in QBism, quantum state assignments are always subjective.  They depend on the degrees of \emph{belief} of the agent using quantum theory, and two agents may assign different quantum states to the system even in the case where they both have access to the same information.  Other interpretations, such as Healy's quantum pragmatism and Rovelli's relational quantum mechanics, argue that quantum state assignments are objective.  

Note that this is \emph{not} the same as the distinction between objective and relativist interpretations.  Relational quantum mechanics is relativist in that the properties (or outcomes) assigned to a system depend on the external system relative to which it is being described. But given a choice of external system, there is an objectively correct quantum state assignment relative to that system.

For present purposes, the question of whether state assignments are objective or subjective is not relevant.  Even if there is an objectively correct state, all Copenhagenishts agree that it is not an \emph{intrinsic} property of an \emph{individual} quantum system.  To quote Bub, the quantum state is not a ``truthmaker of propositions''~\cite{Bub2010}.  Although the quantum state encodes our predictions about quantum experiments, it is not part of the essential nature of the individual system involved in a single run of the experiment.

This implies that, unlike the orthodox interpretation, Copenhagenish interpretations cannot and do not endorse the eigenvalue-eigenstate link, since this link implies that the quantum state is an intrinsic property of a system.  We discuss this further in Section~\ref{sec:orthodox}.

Indeed, as we will discuss further in the next section, the Copenhagenisht need not assign objective intrinsic properties to \emph{any} of the systems that are being described according to quantum mechanics. Therefore, two different quantum state assignments (which may have resulted from two different descriptions of a measurement process) need not be contradictory; they may simply represent the descriptions of two different observers, and the two descriptions need not be unifiable into a single global picture.

\subsubsection{Completeness}\label{Copenhagen:Completeness}

Completeness is the assumption that there is no deeper description of the world to be had than that provided by quantum theory.  Taken together with Universality and Anti-$\psi$-ontology, this has far-reaching implications.  Universality implies that any system can be given a quantum description, Anti-$\psi$-ontology implies that such a description does not assign objective intrinsic properties to the system, and then Completeness implies that we cannot add anything further to that description.  This might seem to imply that systems can have no objective intrinsic properties at all, but this is not quite the case.   It merely implies that, if there are any such properties, then physics cannot provide a description of them when we are describing the system according to quantum mechanics.  In other words, any such properties are \emph{ineffable}\footnote{Ineffable is usually defined as ``incapable of being expressed in words''.  Here we mean that it is literally indescribable in any terms.  In a slogan: the moon may be there when you are not looking at it, but it is fundamentally impossible for you to describe its properties in language, pictures, mathematics, computer code, or anything else.}.  Therefore, any such properties do not provide a hidden variable theory or ontological model, since in those frameworks it is assumed that we can supply a physical theory that explains our experiments in terms of the properties ascribed by the hidden variables or ontic states.

When we are describing a system according to quantum mechanics, the three assumptions imply that there is {\em no} description of the objective intrinsic properties of an individual system to be had.  Therefore, Copenhagenish interpretations must be antirealist to a certain extent; i.e., when we are describing a system according to quantum mechanics, then we cannot also speak meaningfully about its objective intrinsic physical properties.

However, Observers Observe implies that some systems do have physical properties that we would presumably like to be able to talk about; namely the results of observations made on quantum systems.  In order to talk about them, those results must be described classically, which entails that we must split the world into a part we are describing according to quantum mechanics, which includes the system that is being observed, and a part we are describing classically, which includes the measurement outcomes.  Depending on the Copenhagenish interpretation under consideration, these classical properties might be assigned to the observer alone, to the system relative to the observer, or to the interaction between the system and observer.  For simplicity, we will tend to use the shorthand of describing measurement results as properties ``of the observer'', but this is meant to encompass the other two possibilities as well.  What is important is that the measurement results cannot be intrinsic properties of the system alone because the system is being described according to quantum mechanics.

In this sense, Copenhagenish interpretations need not be wholesale antirealist, e.g.\ they need not be antirealist about systems for which classical mechanics provides an adequate account, or about the subject matter of higher-level sciences like biology, but they are antirealist about intrinsic properties of systems that are being described quantum mechanically.

Another consequence of Completeness is that most Copenhagenishts reject counterfactual reasoning, e.g., about what would have happened had I made a different measurement from the one I actually chose, as this would entail the existence of properties that go beyond the predictions of quantum mechanics. Also, any concrete claims about events that are in principle unobservable are typically viewed as meaningless on the same grounds.

Note that the term $\psi$-completeness was defined in the ontological framework to be the conjunction of Completeness and $\psi$-ontology~\cite{Harrigan}.  Now, $\psi$-completeness is a much stronger assumption that is rejected in Copenhagenish interpretations, since they reject $\psi$-ontology.  However, some authors use the term Completeness to mean $\psi$-completeness, i.e.\ they define Completeness to mean that the quantum state specifies all of the physical properties of a system.  We do not think that a different meaning of Completeness is intended by these authors; rather it seems that $\psi$-ontology is taken for granted by them.  Two factors that contribute to this are that the orthodox interpretation, which is $\psi$-complete, is often conflated with the Copenhagen interpretation, and that the best known realist interpretations---Everett/many worlds, de Broglie-Bohm theory, and spontaneous collapse theories---are all $\psi$-ontic.  So it is fairly easy to fall into the trap of believing that, apart from \emph{wholesale} antirealism, $\psi$-ontic interpretations are the only serious possibility\footnote{This impression may be exacerbated by the Pusey-Barrett-Rudolph theorem~\cite{pusey2012reality} and other $\psi$-ontology theorems, but these are all proved in the ontological models framework, which is rejected by Copenhagenishts as well as realist interpretations with exotic ontology, such as Everett/many-worlds, interpretations that involve retrocausality~\cite{RevModPhys.92.021002,PRICE201275,adlam2022two}, or interpretations that involve providing nonclassical causal explanations~\cite{schmid2021unscrambling,ormrod2024quantuminfluenceseventrelativity}.}.  However, as we explain in Section~\ref{sec:orthodox},  Copenhagenish interpretations are not the orthodox interpretation, and they are not wholesale antirealist, so the assumptions of $\psi$-ontology and Completeness should be separated. This can lead to confusion\footnote{For instance, conflating the two led Ref.~\cite{Okon2022,okon2022reassessing} to argue that the conjunction of Completeness, Universality and the objective version of Observers Observe (used in some extended Wigner's friend arguments) are already ruled out by Maudlin's trilemma, but this is not true without an implicit assumption of $\psi$-ontology.}, as some important works use Completeness to mean $\psi$-completeness; for example, Maudlin uses the word ``completeness'' in this way in his well-known formulation of the measurement problem as a trilemma~\cite{maudlin1995three}.

\subsection{What Copenhagenish interpretations are not}\label{Copenhagen:Not}

Since the Copenhagen interpretation is often thought to be the default interpretation of quantum mechanics, it is often conflated with what is found in quantum mechanics textbooks.  Most textbooks on quantum mechanics are not primarily concerned with interpretation, and they usually describe something closer to what we call the \emph{orthodox} interpretation of quantum mechanics, or try to avoid the issue altogether by adopting the ``shut up and calculate'' attitude.  Neither of these views is at all close to the Copenhagen interpretation or its modern descendants.  Many issues that are thought to be problems for Copenhagenish interpretations, or for quantum mechanics per se, such as the measurement problem and Wigner's Friend, are in fact only problems for the orthodox interpretation.  We explore this in~\cref{sec:MMP,sec:OGwigner}.

\subsubsection{The orthodox interpretation}
\label{sec:orthodox}

The key postulate of the orthodox interpretation (also known as the \emph{textbook} or \emph{Dirac-von Neumann} interpretation) is the \emph{eigenvalue-eigenstate link}\footnote{The term was first coined by philosopher Arthur Fine~\cite{Fine1973}.  Gilton~\cite{Gilton2016whence} presents the case that the principle, although unnamed, is present in early textbook accounts of quantum mechanics.}.  This rule specifies intrinsic properties for a quantum system, and, as such, it directly contradicts the Copenhagenish principle (alluded to above and discussed further in Section~\ref{sec:consequence}) that systems being described quantum mechanically should not be assigned intrinsic properties.  It also renders the orthodox interpretation $\psi$-ontic (in fact $\psi$-complete), so it directly contradicts the Copenhagenish principle of Anti-$\psi$-ontology.
  
The eigenvalue-eigenstate link states that an observable, represented by a self-adjoint operator $A$, corresponds to an intrinsic property of a quantum system if and only if the quantum state $\Ket{\psi}$ is an eigenvector of $A$.  Further, the corresponding eigenvalue is the value of the property that obtains.  Finally, echoing Completeness, these are the \emph{only} intrinsic properties of the system\footnote{Apart, perhaps, from things that do not vary as we change the quantum state, such as the dimension of the Hilbert space of the system.}. For example, a particle has an intrinsic property of spin in the $z$-direction when the quantum state is an eigenstate of the Pauli $\sigma_Z$ operator, but, in this case, it does not have an intrinsic property corresponding to spin in the $x$-direction.

In addition, the orthodox interpretation assumes that measurements have a single unique outcome, i.e., it is not a many-worlds theory, and it endorses the textbook postulates of quantum mechanics, including the fact that measurements must generally change the quantum state of the system and that they can be implemented in such a way that the projection postulate holds.

The eigenvalue-eigenstate link renders the orthodox interpretation $\psi$-ontic.  To see why, suppose that the system is assigned the quantum state $\Ket{\psi}$ and consider the observable represented by the rank-1 projector $\Proj{\psi}$.  Now, $\Ket{\psi}$ is an eigenvector of  $\Proj{\psi}$ with eigenvalue $1$, so this observable is an intrinsic property of the system that takes value $1$.  Further, up to an global phase, $\Ket{\psi}$ is the \emph{only} quantum state vector that has eigenvalue $1$ for this observable.  Knowing which rank-1 projector is a property of the system with value $1$ suffices to determine the quantum state of the system, but, by definition, a $\psi$-ontic interpretation is one in which knowing the intrinsic properties of the system determines the quantum state uniquely.  

Further, the orthodox interpretation is $\psi$-complete because once you have specified the quantum state $\Ket{\psi}$, all of the other properties of the system are determined by it.  For any observable represented by an operator $A$, you simply have to check whether $\Ket{\psi}$ is an eigenvector of $A$ and, if so, what the corresponding eigenvalue is.

The intuition behind the eigenvalue-eigenstate link is that if it is possible to determine the outcome of a measurement with probability one, and moreover possible to do so without changing the quantum state of the system, then that is grounds for asserting that the quantity being measured was already a property of the system beforehand.  The lack of change of the quantum state is taken as an indication that the system is not disturbed by the measurement (although this implicitly assumes the $\psi$-complete view).  This intuition is similar to motivation for the Einstein-Podolsky-Rosen (EPR) criterion for reality, except that for EPR, the lack of disturbance of the system is supposed to be enforced by spacelike separation rather than lack of change of the quantum state.   

Copenhagenish interpretations reject the eigenvalue-eigenstate link for several reasons.  First, this link requires that there be a unique, objectively correct, quantum state to assign to a system.  Up to a global phase, each distinct state vector is either an eigenstate of  a different set of observables or has a different eigenvalue for at least one observable.  Thus, either the definite properties or the values of those properties are necessarily different for different state vectors.  But since these are the intrinsic physical properties of the system, they must be uniquely specified, so only one state assignment can be correct, and this violates the Anti-$\psi$-ontology.  This means that the measurement problem poses an extreme difficulty for the orthodox interpretation.  It can avoid inconsistency only by rejecting Universality.  We will discuss this more in Section~\ref{sec:MMP}.  Second, the orthodox interpretation requires the existence of superluminal influences via the EPR argument, whereas a supposed virtue of Copenhagenish interpretations is that they avoid the need for nonlocality. 

Part of the reason that the Copenhagen interpretation became conflated with the orthodox interpretation is that the Copenhagen founders often used ontological and epistemological language interchangeably and inconsistently.  The epistemological counterpart of the eigenstate-eigenvalue link is: the outcome of a measurement of an observable can be predicted with certainty if and only if the state vector assigned to the system is an eigenstate of the corresponding self-adjoint operator, so in this sense, these are the only physical quantities that can be ``known'' with certainty.  This much would be accepted as fact by every physicist, up to minor quibbles about terminology (e.g., whether ``known'' has the implication that the system possessed the physical quantity before the measurement was made, or whether we should rather say that the measurement ``created'' the outcome).  All of this is compatible with the Copenhagenish principles, as well as any other interpretation of quantum mechanics, because it only refers to the operational predictions of the theory, which everybody agrees upon.  

Things only become problematic when we replace talk of ``predictions'' and ``knowledge'' with talk of ``properties'' and ``definite values''.  Then one is dealing with ontology, which is the realm of the eigenvalue-eigenstate link.

In a historical study~\cite{Gilton2016whence}, Gilton traces the increasing significance given to the eigenvalue-eigenstate link in the major quantum mechanics textbooks from Heisenberg to Messiah.  Gilton argues that a version of the link is stated in Heisenberg's book, although not in modern terminology.  The link is stated in Dirac's textbook, although much more explicitly in the second edition than the first.  It is stated in a more mathematical language by von Neumann (whence the name Dirac-von Neumann interpretation).  By the time we get to Messiah, the eigenstate-eigenvalue link is clearly highlighted as an important principle of quantum mechanics, as it appears in italics in a separate paragraph.  Based on this, Gilton argues that the eigenvalue-eigenstate link is part of ``standard'' quantum theory.  Due to Heisenberg's statement, one might even think that it is part of the Copenhagen interpretation.

The problem with Gilton's analysis is that, with the exception of von Neumann, all of the authors Gilton studies state both something similar to the eigenvalue-eigenstate link \emph{and} something similar to its epistemic counterpart.  They do so without noting that there is a distinction, and often in the same section of the textbook.  Gilton takes this as evidence that they intended that \emph{both} principles should hold, but you could just as easily argue that it means they had not clearly understood the difference between the ontological and epistemic principles, and so which one they intend to endorse is ambiguous.  The exception to this is von Neumann, who only states the eigenvalue-eigenstate link in terms of the presence or absence of properties, which is clearly ontological.   

In conclusion, we believe that the orthodox interpretation arose from an ontological reading of a principle that should only have been used as an epistemological principle by the Copenhagen founders.  It was conflated with Copenhagenishism only because the two readings of the principle were not clearly distinguished at the time. (This is good news for the Copenhagenishts, since the orthodox interpretation is either inconsistent or incomplete, as we will see in Section~\ref{sec:MMP}.)

\subsubsection{Shut up and calculate}

David Mermin once stated ``If I were forced to sum up in one sentence what the Copenhagen interpretation says to me, it would be `Shut up and calculate!' ''.
This term (often misattributed to Feynman~\cite{ShutUp}) is meant to invoke the attitude of many practicing physicists who are satisfied with the predictive power granted by quantum theory, and uninterested in the interpretative question of what it teaches us about reality.

It is a mistake to associate this view with Copenhagenish interpretations. Indeed, the ``shut up'' part is an extremely inaccurate description of the works of founders like Bohr and Heisenberg, who wrote voluminously on the subject of interpretation and talked about it at great length.  Bohr famously argued with Schr{\"o}dinger at such length that Schr{\"o}dinger became ill and took to his bed, which Bohr failed to take as a signal to stop talking to him about quantum mechanics.  Modern Copenhagenishts also fail to ``shut up'' in their voluminous works\footnote{One of them famously posts their entire email correspondence on the subject on the internet~\cite{fuchs2001notespaulianideafoundational,fuchs2015strugglesblockuniverse}.}.  Indeed, Mermin later disowned his ``shut up and calculate'' statement, calling it a ``childishly brusque dismissal of such an exquisitely subtle point of view''~\cite{ShutUp}.  He subsequently came up with his own Copenhagenish interpretation~\cite{Mermin1998}, before becoming a latter day QBist~\cite{FuchsMerminSchack}. 

Mermin's conflation of Copenhagenishism with ``shut up and calculate'' is understandable from a historical point of view.  As Kaiser has argued~\cite{Kaiser2022}, prior to the 1940's most quantum mechanics textbooks were published in Europe and contained lengthy discussions of the philosophical implications of the theory, usually along Copenhagen lines (although perhaps with an undue emphasis on the eigenvalue-eigenstate link).  During the second world war, physics took a much more pragmatic and practical turn, partly because the center of gravity of physics moved from Europe to the USA, which has always taken a more practical attitude towards science, and partly due to the implications of the Manhattan project, which suddenly made fundamental physics relevant to urgent political, ethical and technological issues.  Regardless of the reason, in the post-war period, textbooks vastly reduced their discussion of the interpretation of quantum mechanics and work on the foundations of quantum mechanics became taboo in physics departments.  The standard view became that it is pointless to worry about such issues because Bohr, Heisenberg, et. al. had sorted everything out back in the 1930's, so it is easy to see how Copenhagen got conflated with ``we don't worry about it''. 

\subsubsection{Operationalism}

The ``shut up and calculate'' view is a kind of agnosticism about the interpretation of quantum mechanics, underpinned by the claim that differences in interpretation will never make a difference to the predictions of the theory.  However, if this point of view is taken to its logical conclusion then we arrive at strict operationalism or instrumentalism, which claims that it is not the job of physics to give an account of the physical world independent of our experiments.  Instead, concepts like ``preparation'', ``measurement'' and ``transformation'' are taken as primitives and the job of physics is simply to predict the probabilities of measurement outcomes in experiments composed of these primitives.  Given that ``shut up and calculate'' gets conflated with Copenhagen, it is understandable that operationalism does as well.

But Copenhagenish interpretations are \emph{not} strict operationalism.  They do not forbid you from believing that quantum systems have properties independently of whether they are measured in an experiment.  They only assert that you can say very little about those properties when you are describing a system according to quantum mechanics.  Copenhagenishts are also usually realist about the macroscopic world of our experience, even though they do forbid one from giving a straightforwardly realist description of a system that is being treated quantum mechanically.

\subsection{Shifty splits}
\label{sec:consequence}

Here we detail an important consequence of the four postulates of Copenhagenish interpretations.

On first encounter, the two postulates of special relativity seem to contradict each other.  How can the laws of physics be the same in all inertial reference frames if there is also an absolute velocity, which should pick out a preferred frame?  The apparent contradiction only holds if we keep all the other structures of classical physics held fixed.  In particular, revising the Newtonian concepts of space and time allows the postulates to peacefully coexist.  Likewise, the defining postulates of Copenhagenish interpretations initially seem to be in conflict.  How can quantum theory be complete, universal, and have the straightforward existence of measurement outcomes if it denies that the things being described by quantum theory (including observers and measuring devices) can have intrinsic properties? How can state-update rules {\em and} unitary dynamics both be universally applicable, given that the latter describes deterministic and continuous dynamics, while the former describes indeterministic and discontinuous dynamics? Such apparent conflicts can again be resolved by making a radical change to the structure of physics elsewhere, i.e., by rejecting the requirement that the goal of physics is to provide a unique objective account of the properties of the world, independent of observers. Rather, the goal of physics is more focused on epistemology than ontology---more focused on giving a prescription for reasoning about what observers might observe, rather than on giving a description of what actually happens independently of observation (including on the microscale).

In this section, we discuss how this follows from the interplay between and consequences of the four defining postulates. 

On the face of it, you may be surprised that we list Universality as a Copenhagenish principle, given that Bohr insisted on the need to describe our experimental arrangements in the language of classical physics, which necessitates a split (usually called the \emph{Heisenberg cut}) between the quantum system we are investigating and the apparatus that we describe classically.  However, this split is supposed to be movable, so one can give a quantum description of any system in the universe, so long as it is embedded in a larger system, the rest of which is described classically.  The movability of this split has led to it being termed ``the shifty split'' \cite{Bell1990}, but---to be clear---this shiftiness is deliberate.  Normally, if a theory is built on operational principles, one hopes to later give it a realist underpinning in terms of a more fundamental physical theory.  Such is the relationship between equilibrium thermodynamics and statistical mechanics for example.  But the Copenhagen founders thought that quantum theory was a fundamental theory of physics, not an operational expedience to be replaced by something deeper.  Since they viewed it as fundamental, they had little choice but to think that it applies to everything in principle; however, due to their empirical leanings, they still needed something outside the quantum mechanical description to which their empirical records could refer.  The shifty split is thus an attempt to have their cake and eat it too, i.e., for the theory to be fundamental, but also partly operational. 

The cut cannot be moved arbitrarily, however. When you attempt to describe an experiment, there is a highest level and a lowest level at which you can put the cut.
The lowest level arises because if you coherently interfere a degree of freedom which is viewed as being on the classical side of the cut, then you will get the predictions wrong. (Today, we might use decoherence theory~\cite{RevModPhys.75.715} to decide where the lowest level for the cut is.)
The highest level arises because analyzing an experiment requires us to reason about the immediate sensory experiences of observers; as those experiences are only describable in classical terms, we must place those observers on the classical side of the cut. That is, if I want to account for the particular definite outcomes that someone observed in a given run of an experiment, then that is ascribing a property {\em to that observer}, and so I cannot include them in my quantum description.

For each different placement of the cut, you get a different description of a given experiment. The fact that different quantum states are assigned at different levels is not a problem, because Copenhagenishts are Anti-$\psi$-ontologists. All that matters is that the different levels agree on the predictions for experiments that are {\em actually performed}. 
Whether or not it is the case that these different descriptions are consistent, and whether they can be combined in any nontrivial ways, is not immediately clear, and is the subject of debate in the literature. We will elaborate on these questions in the context of some specific examples from Section~\ref{sec:MMP} onwards.

It is quite subtle to reconcile the fact that there are different incompatible descriptions of any given physical experiment.  Different Copenhagenish interpretations disagree on if or how these descriptions can be combined or seen to be consistent. These problems ultimately arise because there is a highest level for the Heisenberg cut, as this prevents one from simply treating {\em all} systems quantum mechanically. However, the existence of a highest level is necessary, both as a consequence of the Copenhagenish idea that all personal experience must ultimately be described classically, and also to ensure consistency of such interpretations. As we will see more explicitly when we discuss variations of Wigner's friend thought experiments, modelling an observer unitarily while also reasoning about the outcomes they observed leads to inconsistencies.
As such, it is useful to expand on what it means to reason about observations.

There are multiple distinct senses in which you might ``reason about the observations of an observer'', all of which are subject to some degree of suspicion in Copenhagenish interpretations:
\begin{enumerate}
    \item You might assert that the observer saw an outcome.
    \item You might assert that the relative frequency of these observed outcomes is distributed according to the Born rule.
    \item You might condition your reasoning on the observer having seen a particular outcome. 
    \item You might apply a state-update rule to the state assigned by an agent to assign them a new quantum state based on which outcome they observed.
\end{enumerate}

This is not to say that when one chooses to model an observer as a quantum system, one must believe that the observer will {\em not} observe an outcome---indeed, to do so would be to reject Observers Observe. It is just that in any argument or calculation about what can be expected to occur in some experiment, you cannot combine facts about the observer's observed outcomes together with facts that follow from treating the observer as a quantum system. The two perspectives are incompatible (or complementary, in the sense of Bohr), and so are not to be combined.

This is awkward from the standpoint of a straightforward realist picture of the world---if you believe that there is an absolute fact about what the observer observed (as in objective Copenhagenish interpretations), and about the frequencies with which each observation occurs, then why should one ever be forbidden from using this knowledge in one's reasoning? But it is perhaps sensible from a Copenhagenish perspective wherein quantum theory is more of an epistemological tool than a picture of the world that extends to the microphysical level. And in a realtivist Copenhagenish interpretation, the impossibility of combining these distinct accounts makes good sense, because the ``facts'' to be combined are not generally both true relative to any single observer.

\section{The measurement problem}
\label{sec:MMP}

There are several aspects to what is normally called ``the measurement problem''.  The first is that the postulates of quantum mechanics refer to the notion of ``measurement'', but they give no precise criterion for when a measurement occurs.  This is apparently problematic because the theory says that an experiment produces data only when a measurement occurs, and the rules for updating the quantum state over time differ depending on whether or not a measurement has occurred.  In the orthodox interpretation, where the quantum state determines which properties exist, differing state assignments imply different properties for the system, so the two rules for updating the state are contradictory.  Therefore, the orthodox interpretation can only be made consistent if we provide a precise criterion for when a measurement has occurred.  This means that the orthodox interpretation cannot endorse Universality, which implies that the interaction between a system and an observer can always be described by a unitary interaction.

In Copenhagenish interpretations, there is no question of dropping Universality, as it is one of the postulates of this class of interpretation.  Hence, a Copenhagenish interpretation cannot endorse any precise criterion for when a measurement has occurred.  The reason that this does not immediately lead to inconsistency with the two different rules for updating a quantum state over time (depending on whether or not a measurement has occurred) is because of Anti-$\psi$-ontology.  Having two different state assignments does not necessarily imply contradictory statements about the properties of the system.

Let us consider the quantum description of a measurement in a bit more detail. We will follow the presentation in Ref.~\cite{schmid2023review}.  Suppose a qubit system, labeled ${\rm S}$, has an associated Hilbert space $\Hilb[S]$ and is prepared in the state $\Ket{+}_{\rm S} = \frac{1}{\sqrt{2}} \left ( \Ket{0}_{\rm S} + \Ket{1}_{\rm S} \right )$.  The system is measured in the computational basis $\left \{ \Ket{0}_{\rm S}, \Ket{1}_{\rm S} \right \}$ in such a way that the projection postulate applies.  According to the measurement postulates of quantum mechanics, the $\Ket{0}_{\rm S}$ outcome will be obtained with probability $1/2$, and if it is obtained, the state is updated to $\Ket{0}_{\rm S}$. Similarly, the $\Ket{1}_{\rm S}$ outcome will be obtained with probability $1/2$, in which case the state is updated to $\Ket{1}_{\rm S}$.  The net result is that after the measurement, the state will be either $\Ket{0}_{\rm S}$ or $\Ket{1}_{\rm S}$, with probability $1/2$ each. 

However, a measurement device is just a physical system, made of atoms like everything else, so according to Universality, we should be able to associate it with a Hilbert space $\Hilb[M]$, so that the interaction between the qubit and the measuring device can be described by a unitary on $\Hilb[S]\otimes\Hilb[M]$ according to the Schr{\"o}dinger equation. Note that $\Hilb[M]$ should be taken to be a large enough subsystem of the universe such that the interaction between the system and the measuring device can always be modeled by a unitary.  If one thinks that environmental decoherence is necessary for the device to register a definite measurement outcome, then a significant portion of the environment beyond the measurement device itself should also be included in $\Hilb[M]$.

Suppose that the measurement device starts in a ready state $\Ket{R}_{\rm M}$, a state in which the measurement device is ready to start a new measurement, e.g., it is calibrated and the pointer needle points to zero.  Note that there will typically be many possible $\Ket{R}_{\rm M}$ states, e.g., because parts of the apparatus may be in thermal contact with the environment.  All that matters is that the measurement device is in one such state, as they will all behave similarly in the measurement interaction.

Assuming that the measurement device works correctly when the system is prepared in either the $\Ket{0}_{\rm S}$ state or the $\Ket{1}_{\rm S}$ state, the unitary evolution $U_{\rm SM}$ describing the measurement must satisfy
\begin{equation}
\label{eq:Usm1}
    \begin{split}
	U_{\rm SM} \Ket{0}_{\rm S}  \Ket{R}_{\rm M} & = \Ket{0}_{\rm S}  \Ket{0}_{\rm M}, \\
	 U_{\rm SM} \Ket{1}_{\rm S}  \Ket{R}_{\rm M} & = \Ket{1}_{\rm S}  \Ket{1}_{\rm M}, 
    \end{split}
\end{equation}
where $\Ket{0}_{\rm M}$ and $\Ket{1}_{\rm M}$ are states in which the measurement device is registering the outcome $0$ and $1$ respectively.  Note that, as with $\Ket{R}_{\rm M}$, there will typically be many possible $\Ket{0}_{\rm M}$ and $\Ket{1}_{\rm M}$ states, but all that matters is that the evolution takes $\Ket{0}_{\rm S}  \Ket{R}_{\rm M}$ and $ \Ket{1}_{\rm S}  \Ket{R}_{\rm M}$ to one such pair of states.  Typically we want to have $\BraKet{0}{1}_{\rm M} \approx 0$, so that the outcome can be determined with good accuracy by measuring the measurement device, but this is actually immaterial for the argument.

For some ways of implementing a measurement, $\Ket{R}_{\rm M}:=\Ket{0}_{\rm M}$, in which case the unitary evolution for the measurement satisfying Eq.~\eqref{eq:Usm1} is the CNOT gate\footnote{For example, suppose the system is an ion in an ion trap that is confined to be in its ground state $\Ket{0}_{\rm S}$ or first excited state $\Ket{1}_{\rm S}$.  We assume that a selection rule prevents the spontaneous transition from $\Ket{1}_{\rm S}$ to $\Ket{0}_{\rm S}$, so that we can reliably encode a qubit in the $\Ket{0}_{\rm S},\Ket{1}_{\rm S}$ subspace.  One way of measuring the energy of the ion is to excite the ion to a higher energy state $\Ket{2}_{\rm S}$ with a laser pulse, tuned so that the transition $\Ket{1}_{\rm S}$ to $\Ket{2}_{\rm S}$ is on resonance, but the transition $\Ket{0}_{\rm S}$ to $\Ket{2}_{\rm S}$ is blocked by a selection rule.  The measurement consists of applying this pulse and then detecting any fluorescence of the ion on a screen.  The state of the screen at the beginning of the measurement, and at the end if the outcome is $\Ket{0}_{\rm S}$, will be the same---a state $\Ket{0}_{\rm M}$ showing no fluorescence on the screen.  However, when the outcome is $\Ket{1}_{\rm S}$ the screen will transition to a state $\Ket{1}_{\rm M}$ showing fluorescence on the screen.  Although this is not the only way of measuring the ion's energy, it is one possible way of doing it, and so any general problems with our understanding of quantum measurement ought to apply to this example in particular.}. From now on, we assume this convention, so Eq.~\eqref{eq:Usm1} becomes
\begin{equation}
    \label{eq:Usm}
        \begin{split}
        U_{\rm SM} \Ket{0}_{\rm S}  \Ket{0}_{\rm M} & = \Ket{0}_{\rm S}  \Ket{0}_{\rm M}, \\
         U_{\rm SM} \Ket{1}_{\rm S}  \Ket{0}_{\rm M} & = \Ket{1}_{\rm S}  \Ket{1}_{\rm M}, 
        \end{split}
    \end{equation}

Now consider the case where the qubit is prepared in the $\Ket{+}_{\rm S}$ state, and thus the joint state of the qubit and the measuring device is $\Ket{+}_{\rm S} \Ket{0}_{\rm M}$. We apply $U_{\rm SM}$ to $\Ket{+}_{\rm S} \Ket{0}_{\rm M}$ and obtain, by linearity, the joint state after the measurement
\begin{align}
    U_{\rm SM} \Ket{+}_{\rm S} \Ket{0}_{\rm M} &= \frac{1}{\sqrt{2}} \left ( U_{\rm SM} \Ket{0}_{\rm S}  \Ket{0}_{\rm M} + U_{\rm SM} \Ket{1}_{\rm S}  \Ket{0}_{\rm M} \right ) \notag \\
                              & = \frac{1}{\sqrt{2}} \left ( \Ket{0}_{\rm S}  \Ket{0}_{\rm M} + \Ket{1}_{\rm M}  \Ket{1}_{\rm M}\right ) =: \Ket{\Phi^{+}}_{\rm SM}. \label{entangSM}
\end{align}

This is clearly not the same quantum state as one obtains in the first description of the measurement process (either $\Ket{0}_{\rm S}$ or $\Ket{1}_{\rm S}$).  From the point of view of the orthodox interpretation, these two descriptions are in flat-out contradiction since the orthodox interpretation endorses the eigenvalue-eigenstate link. This implies that, if the state of the system is either $\Ket{0}_{\rm S}$ or $\Ket{1}_{\rm S}$, then the system has a property corresponding to the projector $\Proj{0}_{\rm S}$, which takes either value $1$ or $0$ depending on whether the state is $\Ket{0}_{\rm S}$ or $\Ket{1}_{\rm S}$.  If the state of the system and the measurement device is $\Ket{\Phi^{+}}_{\rm SM}$, then no such property exists because $\Ket{\Phi^{+}}_{\rm SM}$ is not an eigenstate of $\Proj{0}_{\rm S} \otimes  I_{\rm M}$.  From the orthodox view, then, one of these descriptions must be right and the other wrong.  There must be a fact of the matter about whether a measurement, and hence a state collapse, has occurred. So the orthodox interpretation is inconsistent with Universality, and it is incomplete unless it gives a definite prescription for when a measurement has occurred.

However, this reasoning does not apply to Copenhagenish interpretations.  The two different descriptions can both be correct, but applicable in different circumstances, depending on where you choose to place the shifty split.  Which description you should use depends on whether you choose to treat the measurement device according to quantum mechanics or not. Because of  Anti-$\psi$-ontology, and the fact that Copenhagenishts reject the idea that state assignments imply {\em anything} about the intrinsic properties of the system, having two different quantum state assignments in two different descriptions of the experiment is not a contradiction.  Given that counterfactual reasoning about alternative possible measurements is precluded in Copenhagenish interpretations, the only thing that matters is whether the two descriptions agree on the predictions for measurements that are \emph{actually performed}.

For example, suppose that we want to know what an observer, who is external to the measuring device, will observe about the outcome of the measurement described above.  
%Both descriptions view this observer as classical. 
To reason about the outcomes of this observer, we must place the observer on the classical side of the cut; however, there remains a choice about whether to place the measurement device $M$ on the classical or quantum side of the cut.
If the split is placed between the system and the measuring device, the collapse is applied as soon as the measurement device interacts with the system.  The observer simply sees the outcome already registered on the measuring device, the probability that they see each outcome is $1/2$, and afterwards they will assign the state $\Ket{0}_{\rm S}$ or $\Ket{1}_{\rm S}$ to the system, as they now know which of the states the system collapsed to in its earlier interaction with the measuring device.  On the other hand, if we put the split between the measurement device and the observer, then the collapse is only implemented when the observer looks at the measurement device. However, they will still see each outcome with equal probability, since both terms in the entangled state in Eq.~\eqref{entangSM} have equal amplitude of $\frac{1}{\sqrt{2}}$. Moreover, this will collapse the state of the system and measuring device to $\Ket{0}_{\rm S}\Ket{0}_{\rm M}$ or $\Ket{1}_{\rm S}\Ket{1}_{\rm M}$ afterwards, at which point the descriptions for either placement of the cut assign the same state to the system, and so both descriptions would make the same predictions about any later measurements on the system.

From this we see that the split can be placed at a range of \emph{levels}, where a higher level description places more of the world on the quantum side of the split, without necessarily leading to different predictions for actually performed experiments.

\section{Wigner's friend} 
\label{sec:OGwigner}

The Wigner's friend thought experiment~\cite{Wigner1995} can be thought of as a variant of the Schr{\"o}dinger's cat experiment where a human observer (the Friend) takes the place of the cat, or as a dramatization of the measurement problem where the Friend plays the role of the measurement device.  The Friend, system, and measuring device are sealed in a box, which is an isolated system.  We take the ``Friend'' system ${\rm F}$ to consist of the Friend, the measuring device, and enough of the environment that the interaction between the system ${\rm S}$ and the Friend ${\rm F}$ can be modelled by a unitary interaction from Wigner's point of view.  The friend performs a computational basis measurement on the system, which is prepared in the initial state $\Ket{+}_{\rm S} = \frac{1}{\sqrt{2}} \left ( \Ket{0}_{\rm S} + \Ket{1}_{\rm S} \right )$.  Just as in the measurement problem, we will model this experiment in two ways: one that places the split between $\rm S$ and $\rm F$, and another that places the split between $\rm SF$ and Wigner. We will refer to these as ``the Friend's point of view'' and as ``Wigner's point of view'', respectively, but it should be understood that this is just shorthand (since any agent can reason in the same manner as any other agent). We again follow the presentation in Ref.~\cite{schmid2023review}.
%, with a focus on their implication for Copenhagenish interpretations. \blk

The Friend observes an outcome, based on which they update the state using the projection postulate. If they observe the $0$ outcome, then they update the state of the system to $\Ket{0}_{\rm S}$; if they observe the $1$ outcome, then they update the state of the system to $\Ket{1}_{\rm S}$.  Each possibility is equally likely.

From Wigner's perspective, the Friend is a quantum system, and the system and Friend together form an isolated system that evolves unitarily. Wigner models his Friend in the same way as we modeled the measurement device in the previous section. Let  $\Ket{0}_{\rm F}$ and $\Ket{1}_{\rm F}$ denote the (coarse-grained) quantum states in which the Friend has observed the outcome $0$ or $1$, respectively. Like before, we let $\Ket{0}_F$ also be the ready state of the Friend before the measurement. Then, the unitary evolution $U_{\rm SF}$ describing the measurement must satisfy
\begin{align*}
	U_{\rm SF} \Ket{0}_{\rm S}  \Ket{0}_{\rm F} & = \Ket{0}_{\rm S}  \Ket{0}_{\rm F}, & U_{\rm SF} \Ket{1}_{\rm S}  \Ket{0}_{\rm F} & = \Ket{1}_{\rm S}  \Ket{1}_{\rm F}.
\end{align*}

Wigner models the Friend's measurement by applying  $U_{\rm SF}$ to the initial state $\Ket{+}_{\rm S} \Ket{0}_{\rm F}$, obtaining one of the Bell states
\begin{align}
    U_{\rm SF} \Ket{+}_{\rm S} \Ket{0}_{\rm F} & = \frac{1}{\sqrt{2}} \left ( \Ket{0}_{\rm S}  \Ket{0}_{\rm F} + \Ket{1}_{\rm S}  \Ket{1}_{\rm F}\right ) \nonumber \\
    & = \Ket{\Phi^{+}}_{\rm SF}.
\end{align}
Since Wigner is describing the state of another observer, he is often referred to as a \emph{superobserver}. This term will be even more appropriate in later sections, where we imagine that superobservers also have the extreme technological capability to apply coherent quantum operations to macroscopic systems including other observers.

From the point of view of the orthodox interpretation, this implies a contradiction in exactly the same way as the standard measurement problem of the previous section.
Also as in Sec.~\ref{sec:MMP}, there is no contradiction within a Copenhagenish interpretation: since Copenhagenishts reject $\psi$-ontology, the different state assignments for the system made in the two different perspectives do not reflect different states of reality, just different choices of where to place the shifty split.

The two can easily come into agreement; for example, if Wigner subsequently opens the box and asks his Friend which outcome they saw.  Both Wigner and his Friend are modeling Wigner classically, so at this point, Wigner will collapse the state of the Friend and the system, and they will agree on the final state of the system.

While the box was closed, Wigner assigned a quantum state to the system and Friend, which is something that is impossible to even formulate in the Friend's description. If the Friend wants to reason about their own observations, they necessarily must place themself on the classical side of the split, so the Friend cannot adopt this description. But for the purposes of predicting what the Friend will say they saw after Wigner opens the box and asks, Wigner could adopt the Friend's description, which indeed leads to the same prediction about what he will see when he opens the box. If he does so, then the two will make the same predictions about subsequent measurements made on the system as well.

In general, an observer, Alice, will have a range of possible levels at which they can place the split without changing the predictions for the actually performed experiment.  If they want to reason about some observer's outcomes (be it themself or some other observers, Bob, Charlie, etc.), then they must necessarily put those observers on the classical side of the split.
Just in front of these observers is the highest possible level at which Alice can place the split.  On the other hand, if Alice is going to do an experiment that checks for coherent superpositions of some set of systems, then those systems must be placed on the quantum side of the split, otherwise Alice's predictions will be wrong.  This is the lowest possible level at which Alice can place the split.  Any level between these two will make the same predictions for the actually performed experiments\footnote{In realistic experiments, there may be small differences between the theoretical predictions made at different levels.  Provided these are below the threshold of experimental precision we can include levels that make slightly different predictions within the allowed range.  This will lead to some fuzziness about the location of the lowest and highest levels of the range.}.

We say that two observers, Alice and Bob, have \emph{range compatibility} if they can reach \emph{level agreement}.  This occurs when Alice's highest possible level is higher than Bob's lowest possible level and Bob's highest possible level is higher than Alice's lowest possible level.  In this case, they can agree upon a level that is in the allowed range for both of them.  This is the case in the Wigner's friend experiment.  Although Wigner might start off with his Friend on the quantum side of the split, putting the Friend on the classical side instead will not change Wigner's predictions, so Wigner and his friend can mutually decide to adopt this description for the sake of reaching level agreement.

On the other hand, there is \emph{range conflict} if Alice and Bob cannot reach level agreement.  This happens if Alice's lowest possible level is higher than Bob's highest possible level or vice versa. If range conflicts never occur, then Alice and Bob will never make different predictions\footnote{Since the highest and lowest levels for a given observer are slightly fuzzy, it may not be possible to make a sharp distinction between range conflict and compatibility when the lowest level of one agent is close to the highest level of another.  But it will be clear that there is compatibility when there is a large amount of overlap and conflict when the highest level of one observer is much lower than the lowest level of another.}.  However, the next two experiments show that range conflicts do in fact occur in some experiments.

\begin{figure}[htb!]
    \centering
    \includegraphics[width=0.4\textwidth]{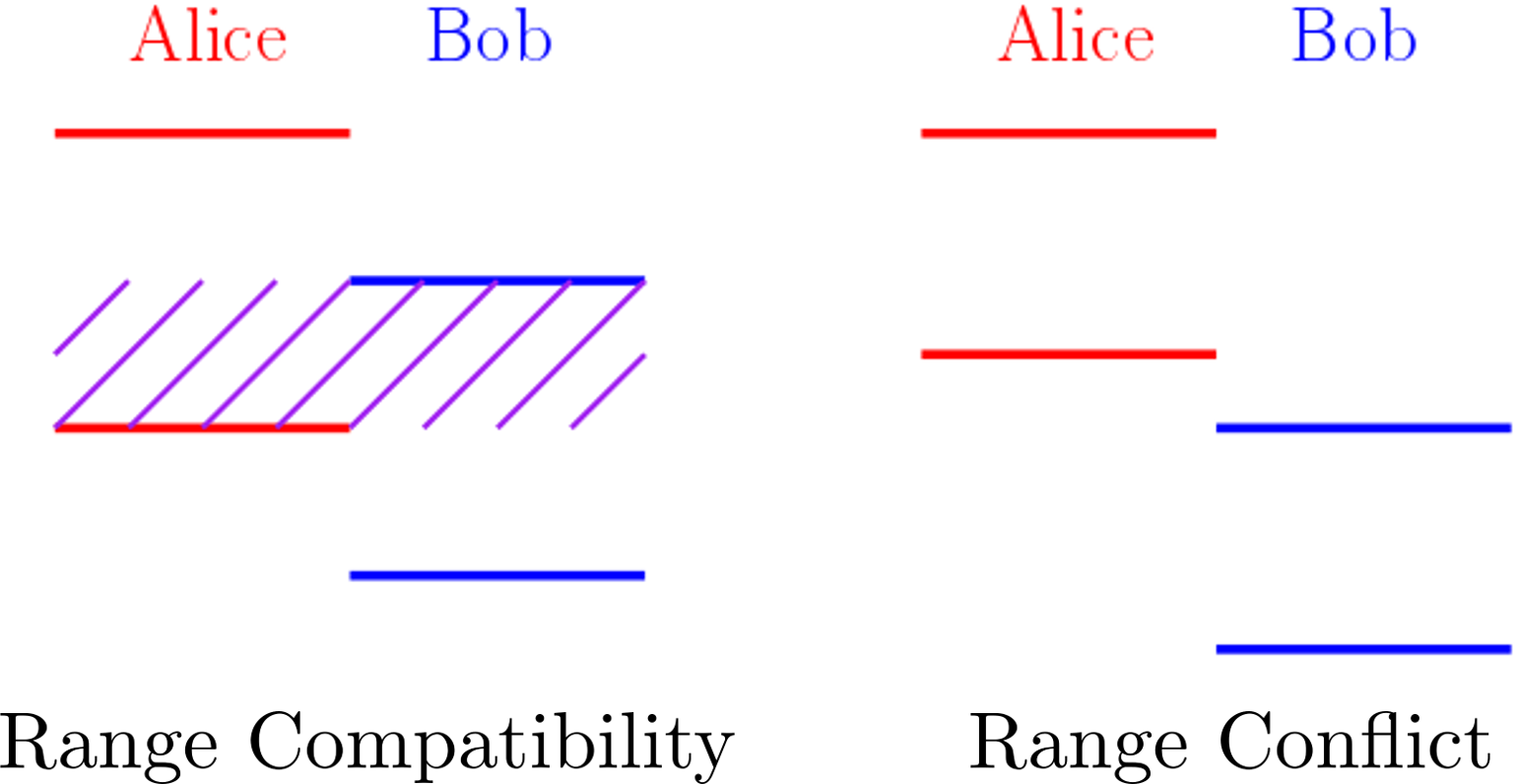}
    \vspace{0.3cm}
    \caption{Illustration of range compatibility and range conflict.}
    \label{range}
\end{figure}

\subsection{Wigner's enemy}
\label{sec:enemy}

Thought experiments of the sorts described in this and in the next subsection were first introduced by Deutsch~\cite{deutschQuantum1985}.

In the Wigner's Friend experiment, Wigner and his Friend have range compatibility, so they can always agree on a common location for the shifty split that is compatible with both of their predictions for the actually performed experiment.  Things would be simple for the Copenhagenisht if every set of observers always had range compatibility.  However, there are situations in which range conflicts necessarily occur.

Suppose Wigner is a superobserver with the (extreme) technological capability to apply unitary operations to the Friend and the system while they are inside the box. For example, he could apply $U_{\rm SF}^{\dagger}$, the inverse of the unitary $U_{\rm SF}$, on the joint system of the Friend and the system ${\rm S}$, effectively reversing (or \enquote{undoing}) the Friend's measurement. According to Wigner, undoing the measurement returns the state of the Friend and system ${\rm S}$ to their initial state, as
\begin{align} \label{eq:Wign}
    U_{\rm SF}^{\dagger}\Ket{\Phi^{+}}_{\rm SF}=\Ket{+}_{\rm S}\Ket{0}_{\rm F}.
\end{align}
In this description, neither Wigner nor the Friend has any knowledge of what the Friend's outcome was after Wigner's undoing operation.
If Wigner were to subsequently open the box and make a measurement on the system in the $\Ket{\pm}_{\rm S}$ basis, he would obtain the $+$ outcome with certainty.  We call this experiment \emph{Wigner's Enemy}, since friends do not recohere friends.

Inside the box, before Wigner's application of $U_{\rm SF}^{\dagger}$, the Friend is aware of the outcome of their measurement, by Observers Observe.  If they want to take account of the outcome they have seen, then they necessarily have to place the split before themselves.  They will collapse the quantum state of the system to either $\Ket{0}_{\rm S}$ or $\Ket{1}_{\rm S}$ depending on which outcome they saw in the measurement.  We suppose that the Friend knows that Wigner is going to implement a procedure on them that Wigner describes as 
$U_{\rm SF}^{\dagger}$, so how should the Friend update the state they assign to the system when this happens?  The difficulty is that the unitary  $U_{\rm SF}^{\dagger}$ is a quantum interaction between the system and Friend, but the friend has no quantum description of themselves.  So it appears that they cannot say anything about how the system changes during this interaction.  But this is not entirely true.

It is permissible for the Friend to neglect their own measurement outcome (that is, to not reason about it) and to reason about Wigner's undoing exactly as Wigner would, placing the cut above themself and so predicting that Wigner will obtain the $\Ket{+}_{\rm S}$ outcome with certainty. Alternatively, the Friend can reason about their own measurement outcome, in which case they must forgo all reasoning about Wigner's undoing, as we just established. 
However, what the Friend {\em cannot} do is combine their reasoning in the two different descriptions into a single picture. On the objective Copenhagenish point of view, this is because Wigner's outcome is ineffable when the split is placed before the Friend, and the Friend's outcome is ineffable when the split is placed above the Friend.  So there is no point of view from which we can combine our reasoning about the two descriptions.  This is not to say that Wigner's outcome and the Friend's outcome do not both exist. 
Observers Observe states that they do. In an objective Copenhagenish interpretation, both facts are true in an absolute sense, and for all observers, but there is an epistemological preclusion on reasoning about them together.\footnote{\label{ft:jointfreq} If both Wigner's outcome and the Friend's outcome exist in an absolute sense, then after many runs of the experiment, there are facts about the joint relative frequencies for both observations. Then why don't they immediately determine the probabilities that the observers ought to assign, which would then determine how they should reason about the observations?  This only applies to a strict frequentist account of probability. On any other account, the actual frequencies are related to the probabilities, but do not completely determine them.  For example, on a subjective Bayesian account, probabilities represent the subjective degrees of belief of an observer, which are updated by conditioning on the observed data.  Provided the prior distribution is generic, and there are a large number of data samples, then the conditioned probabilities will be close to the relative frequencies.  However, if the observer necessarily cannot become aware of some of the measurement outcomes, then they cannot condition on them, and so the conditioned probabilities will not reflect their relative frequencies. For example, suppose an experiment involves observing two variables, $A$ and $B$.  The objective version of observers-observe requires that definite values of $A$ and $B$ are obtained in each run of the experiment and hence that a relative frequency distribution $P(A,B)$ exists in nature.  However, suppose that Alice observes the values of $A$ and Bob observes the values of a $B$ but, by virtue of the way the experiment is conducted, Alice can never become aware of the values of $B$ and Bob can never become aware of the values of $A$.  Then, Alice's marginal probability distribution for $A$ should converge to $P(A)$ over time and Bob's marginal for $B$ should converge to $P(B)$, but nobody's joint probability distribution need converge to $P(A,B)$.  Since the correlations in $P(A,B)$ are unobservable, they may (with no empirical consequences) differ from what one would expect them to be from a naive application of some physical theory.  From an epistemological point of view the marginal $P(A)$ constrains how we should reason about Alice's observations and $P(B)$ constrains how we should reason about Bob's, but the correlational structure of $P(A,B)$ has no such import.  Although this point is easiest to make for a Bayesian approach to probability, it is not limited to this context, since the direct connection between relative frequency and probability is severed in almost all seriously advocated accounts of probability.} On the relativist view of Observers Observe, both facts are true in one or another perspective, but there is no single perspective from which they are both true, and this relativism is what enforces the epistemological preclusion.

From some perspectives (such as hard-headed realism), these restrictions on reasoning may seem overly convoluted and suspect.  What would go wrong if we were to remove the restriction on observers assigning quantum states to themselves while also reasoning about their own measurement outcomes?  If this were allowed then, presumably, before Wigner's undoing of the measurement, the Friend ought to assign either the state $\Ket{0}_{\rm S}\Ket{0}_{\rm F}$ or the state $\Ket{1}_{\rm S}\Ket{1}_{\rm F}$ to the system and themselves, depending on whether they observed the outcome $0$ or $1$.   Consequently, after the application of the unitary, the Friend would assign
\begin{align}\label{eq:EnemyF1}
    U_{\rm SF}^{\dagger}\Ket{0}_{\rm S}\Ket{0}_{\rm F}=\Ket{0}_{\rm S}\Ket{0}_{\rm F}
\end{align}
or
\begin{align}\label{eq:EnemyF2}
    U_{\rm SF}^{\dagger}\Ket{1}_{\rm S}\Ket{1}_{\rm F}=\Ket{1}_{\rm S}\Ket{0}_{\rm F}. 
\end{align}
At this point both Wigner and his friend agree that the system is uncorrelated with the Friend, but they disagree on the state of the system.  If Wigner were to open the box and make a measurement on the system in the $\Ket{\pm}_{\rm S}$ basis, 
%with the Friend observing every step of the process, 
then the two would make different predictions for that measurement.  Wigner would predict that the outcome will be $\Ket{+}_{\rm S}$ with certainty and the Friend would predict that each outcome will occur with probability $\frac{1}{2}$.

Note that, unlike in our treatment of the shifty split in the measurement problem, Wigner and his friend would here make different predictions for an \emph{actually performed} experiment, and they would not have a level disagreement at the time the measurement is performed. They could both put the split between system and Friend for the purposes of analyzing Wigner's measurement.  Their disagreement on the predictions could also persist for an arbitrarily long time, since Wigner could delay performing his measurement for as long as he wants.  So, although the difference of predictions arises because of the range conflict that happened when the friend was inside the box, it would persist even after the conflict was resolved.

How can this be resolved from a Copenhagenish point of view?  First, one might observe that a difference of probabilistic predictions for an actually performed experiment still does not constitute a contradiction provided one takes a sufficiently liberal view of the nature of probabilities (see again  \cref{ft:jointfreq}). For example, on a subjective Bayesian account, there are never any objective facts about what probability must be assigned to an event, so observers may make arbitrarily different probability assignments.   
However, a better response (as noted earlier) is to point out that this argument is illegal from the Copenhagenish point of view, which does not allow the friend to reason about their own measurement outcome while also describing their perception of it according to quantum mechanics at the same time.

\subsection{Wigner's stalkee and penpal}
\label{sec:stalkee}

Next, imagine the following alternative to the Wigner's enemy experiment.  Suppose that the Friend measures the system in the $\Ket{0}_{\rm S}$, $\Ket{1}_{\rm S}$ basis as before, which, according to Wigner, leaves the joint system comprised of the Friend and system  in the entangled state $\Ket{\Phi^{+}}_{\rm SF}$. Now imagine that, instead of undoing the measurement and then measuring $\rm S$ in the $\pm$ basis, Wigner performs a measurement on $\rm S$ and $\rm F$ in a basis containing the projector onto the entangled state $\Ket{\Phi^{+}}_{\rm SF}$. As Wigner's state assignment for the system $\rm SF$ is $\Ket{\Phi^{+}}_{\rm SF}$, he predicts that he will obtain this outcome with certainty. 

Wigner can implement this measurement in a minimally disturbing manner, so that he applies the projection postulate and so his state assignment $\Ket{\Phi^{+}}_{\rm SF}$ is unchanged by the measurement. It might seem natural to think that if Wigner's state assignment has not changed then he has not changed anything about the $\rm SF$ system.  Then, if the Friend's measurement outcome existed before Wigner's measurement, it should still exist, unperturbed, afterwards.  We call this experiment ``Wigner's stalkee'' because Wigner is attempting to keep track of the state of his Friend without doing anything that would alert the Friend to his presence, so he is effectively stalking his Friend.

The difference between this experiment and Wigner's enemy is that, because the Friend's measurement is never undone, both the Friend's outcome and Wigner's exist at the same time.  You might be willing to accept restrictions on what can be said about outcomes that never coexist (like the two outcomes in the Wigner's enemy scenario), but a bit more uneasy about doing this for things that exist at the same time.

In addition to coexisting, it would seem that we can even arrange the experiment so that a single agent has access to both outcomes at once. We simply have
Wigner write the outcome of his measurement on a piece of paper and hand it to the Friend through a hole in the box that is small enough and open for a short enough time that no other interaction between the inside and outside of the box can occur\footnote{The genre of thought experiment in which Wigner and his Friend exchange notes to each other through holes in the box should be called ``Wigner's penpal''. For those who are fed up with the cutesy names for thought experiments, rest assured that this is the last one. Note that the version of Wigner's penpal described here is different from the one described by Deutsch~\cite{deutschQuantum1985}, where the Friend passes a note to Wigner saying ``I observed an outcome''. }.
With this modification, it seems that the Friend could now become aware of both their own measurement outcome and Wigner's at the same time.  Since the Friend is allowed to reason from Wigner's point of view, presumably he agrees with Wigner that the only possible thing that can be written on the paper is, ``I got the $\Phi^+$ outcome.''  The friend is also aware of their own outcome, $0$ or $1$.  We could artificially impose the restriction that the Friend is not allowed to reason about these outcomes together, but that is difficult to make sense of when both outcomes are simultaneously available in the Friends consciousness awareness\footnote{OK, we lied:  ``Wigner's Friend's Multiple Personality Disorder'' anyone?}.

However, this reasoning is a little too quick.  To argue that the Friend's outcome exists (and is unperturbed by Wigner's measurement), we relied upon the fact that in Wigner's description, Wigner's state assignment was unchanged by Wigner's measurement, which leads to the intuition that \emph{nothing} has changed.  But this reasoning is not warranted under Anti-$\psi$-ontology.  The state is not an intrinsic property of the system, nor does it determine any instrinsic properties that the system does have.  The intuition behind this reasoning really stems from the orthodox interpretation, and is not consistent with Copenhagenishism.  Therefore, we cannot say that the Friend's outcome is unperturbed by Wigner's measurement.  In fact, any reasoning about the Friend's measurement outcome requires us to adopt a description where the split is placed before the Friend, and in that description there is no joint state of $\rm SF$ that can be said to remain undisturbed.

Wigner's measurement may or may not have changed the Friend's measurement outcome, but the question of which is the case has no answer in the Copenhagenish point of view.  It is ineffable, because there is no single description that includes both the Friend's outcome and Wigner's measurement.

So, providing one understands the Copenhagenish attitude on the meaning of the quantum state, there is in the end not much difference between the Copenhagenish responses to Wigner's enemy and Wigner's stalkee/penpal. 

\section{Conclusion}

In this work, we proposed a principled characterization of Copenhagenish interpretations of quantum theory, identifying four core postulates---Observers Observe, Universality, Anti-$\psi$-ontology, and Completeness---that capture the commitments common to a broad family of views inspired by the Copenhagen tradition. While proponents of interpretations in this family often emphasize their differences, our aim was to distill the shared assumptions that distinguish Copenhagenish approaches from other competing interpretations.

We then examined several Wigner’s friend-type scenarios---Wigner’s friend, enemy, stalkee, and pen pal. These demonstrate a core feature of the Copenhagenish interpretation---that it often provides multiple complementary descriptions of an experiment, where each description is valid, but not able to be combined with the other descriptions into a single description. To remain consistent, a Copenhagenisht must impose epistemological constraints on combining certain types of information. While these constraints can be imposed in both objective and relativist Copenhagenish interpretations, the advantage of relativist approaches is that they ground the epistemological constraints in the metaphysics of the interpretation: the complementary descriptions cannot be combined into one description because they are not {\em both true} in any one world. Of course, this added simplicity comes at a steep metaphysical price, as these interpretations hold that different observers live in distinct, fragmented realities rather than a single, shared, absolute reality.

These analyses illustrate how Wigner's friend-type arguments engage more meaningfully with the commitments of Copenhagenish views than do many standard no-go theorems in quantum foundations. We hope that our analyses in these simple scenarios will serve as a starting point for a careful exploration of the implications of Copenhagenish views in more complex scenarios, such as extended Wigner's friend scenarios.

\tocless\section{Acknowledgements} This work was supported by Perimeter Institute for Theoretical Physics. Research at Perimeter Institute is supported in part by the Government of Canada through the Department of Innovation, Science and Economic Development and by the Province of Ontario through the Ministry of Colleges and Universities. YY was also supported by the Natural Sciences and Engineering Research Council of Canada (Grant No. RGPIN-2024-04419). ML was supported by Grant 63209 from the John Templeton Foundation.  The opinions expressed in this publication are those of the authors and do not necessarily reflect the views of the John Templeton Foundation.

% \bibliographystyle{IEEEtran}
% \bibliographystyle{apsrev4-2-wolfe}
% \setlength{\bibsep}{3pt plus 3pt minus 2pt}
%\nocite{apsrev42Control}
\bibliography{bib.bib}

%\appendix

\end{document}